\documentclass[aps,prl,tightenlines,twocolumn,amsmath,amssymb]{revtex4}
\usepackage{graphicx}
\usepackage{color}
\usepackage{dcolumn}% Align table columns on decimal point
\usepackage{bm}% bold math
\usepackage{bbm}
\usepackage{subfigure}
\usepackage{pxfonts}
\usepackage{epsfig}

\def\lsim{\mathrel{\rlap{\lower4pt\hbox{\hskip1pt$\sim$}}
    \raise1pt\hbox{$<$}}}                % less than or approx. symbol
\def\gsim{\mathrel{\rlap{\lower4pt\hbox{\hskip1pt$\sim$}}
    \raise1pt\hbox{$>$}}}                % greater than or approx. symbol

\newcommand{\vev}[1]{\langle #1 \rangle}

\newcommand{\Luv}{{\Lambda_{\rm UV}}}
\newcommand{\Lir}{{\Lambda_{\rm IR}}}

\newcommand{\be}{\begin{equation}}
\newcommand{\ee}{\end{equation}}
\newcommand{\bea}{\begin{equation}}
\newcommand{\eea}{\end{equation}}
\newcommand{\nref}[1]{~(\ref{#1})}

\newcommand{\bmp}{\noindent\begin{minipage}{16cm}}
\newcommand{\emp}{\end{minipage}\vskip 7mm} % 7mm untightened

\definecolor{rossoCP3}{cmyk}{0,.88,.77,.40}

\begin{document}
%%%%%%%%%%%%%%%%%%%%%%%%%%%%%%%%%%%%%%%%%%%%%%%%%%%%%%%%%%%%%%%%%%%%%%%%%%%
\title{\Large  \color{rossoCP3} The Mirage of the Fermi Scale %\\ (Untechnicolor )\\
}
\author{Oleg {\sc Antipin}}
\email{antipin@cp3.dias.sdu.dk}
\affiliation{{\color{rossoCP3} {CP}$^{ \bf 3}${-Origins}} \& the Danish Institute for Advanced Study {\color{rossoCP3}\rm{Danish IAS}},  University of Southern Denmark, Campusvej 55, DK-5230 Odense M, Denmark.}%\affiliation{{ CP}$^{ \bf 3}${-Origins}, 
%IFK \& IMADA, University of Southern Denmark, 
%Campusvej 55, DK-5230 Odense M, Denmark.}
\author{Francesco {\sc Sannino}}
\email{sannino@cp3.dias.sdu.dk} 
\affiliation{{\color{rossoCP3} {CP}$^{ \bf 3}${-Origins}} \& the Danish Institute for Advanced Study {\color{rossoCP3}\rm{Danish IAS}},  University of Southern Denmark, Campusvej 55, DK-5230 Odense M, Denmark.}
\author{Kimmo {\sc Tuominen}}
\email{kimmo.i.tuominen@jyu.fi}
\affiliation{Department of Physics, P.O.Box 35, FIN-40014 Jyv\"askyl\"a University, Finland \\ Helsinki Institute of Physics, P.O.Box 64, FIN-00014, Helsinki University, Finland \\
Department of Physics and Astronomy, University of Southampton, Southampton, SO17 1BJ, UK }

%%%%%%%%%%%%%%%%%%%%%%%%%%%%%%%%%%%%%%%%%%%%%%%%%%%%%%%%%%%%%%%%%%%%%%%%%%%%%%%%%%%%%%%%
 \begin{abstract}
The discovery of a light Higgs boson at LHC may be suggesting that we need to revise our model building paradigms to understand the origin of the weak scale. We explore the possibility that the Fermi scale is not fundamental but rather a derived one, i.e. a low energy mirage. We show that this scenario emerges in a very natural way in models previously used to break the electroweak symmetry dynamically and suggest a simple dynamical framework for this idea. In our model the electroweak scale results  from the interplay between two very high energy scales, one typically of the order of  $\Luv\sim 10^{10}$ GeV  and the other around $M_{\rm U} \sim 10^{16}$ GeV, although other values are also possible.
\\[.1cm]
{\footnotesize  \it Preprint: CP$^3$-Origins-2013-23 DNRF90 \& DIAS-2013-23}
\end{abstract}
\maketitle
%%

%\section{Introduction}
It is widely believed that the Fermi scale $F_{\rm{weak}} \simeq 246$~GeV is a fundamental one. This assumption has driven model builders efforts for the past decades. This scale underlies the masses of the weak gauge bosons via the time-honored relation
\begin{equation} 
2\,m_{W} = g \, F_{\rm{weak}}\ ,
\end{equation}
where $m_W$ is the mass of the $W$-boson and $g$ is the weak coupling constant. If we neglect  Quantum Chromodynamics (QCD), every other scale in the Standard model (SM) is related to $F_{\rm{weak}}$. The QCD scale arises purely from interactions and must be fit to experiments, for example, to the proton mass. 
By copying QCD one can naturalize the Fermi scale by replacing the SM Higgs sector with a new strongly coupled theory, and claiming that this {\em technicolor} dynamics generates the Fermi scale in analogy with $\Lambda_{\rm{QCD}}$ of ordinary strong interaction.

The discovery of the Higgs particle \cite{:2012gk,:2012gu}, and confirmation of its properties,
has impeded several model (in)dependent studies of the properties of the observed scalar 
\cite{Carmi:2012yp,Espinosa:2012ir,Giardino:2012ww,Alanne:2013dra}. An interesting further implication 
of the Higgs mass value $M_h\approx 125$ GeV arises from the stability analysis of the Higgs potential: Assuming no new physics beyond the SM, the Higgs potential flattens out and the quartic coupling becomes negative around $\Luv\sim 10^{10}$ GeV.  In other words, if no new physics exists 
beyond the SM, we might live in a metastable universe  \cite{Degrassi:2012ry,Masina:2012tz,Antipin:2013sga}. 
%We stress that 
The correct implementation of the Weyl consistency conditions for the perturbative renormalisation group computations needed to determine the SM vacuum stability appeared in \cite{Antipin:2013sga}. 
Despite all these developments,  we still face the puzzle: What is the origin of the electroweak scale {\em{per se}}?  
 
One logical possibility is that the Fermi scale is not a fundamental one but derived via an interplay of higher scales; de facto a low energy mirage. We will explore this possibility within natural extensions of the SM, constructed as a direct generalization of the composite Higgs and Unparticle \cite{Georgi:2007ek,Georgi:2007si} scenario introduced first in \cite{Sannino:2008nv}. The scenario which we consider is reminiscent of the walking (extended) technicolor: We consider a strongly interacting gauge theory whose matter content is tuned so that the theory lies inside the conformal window. In other words the long distance behavior of the theory is governed by an infrared fixed point at scales 
$\mu\le\Luv$. The strongly interacting sector is coupled to the SM matter fields by extended gauge interactions, which are broken at high scale $M_{\rm{U}}$. What is essential in our case is that
these extended technicolor interactions provide explicit breaking of the scale invariance and may perturb the theory slightly away from the conformal window. We will show that this breaking, when relevant below the scale $\Luv$, leads to the emergence of an infrared scale which we identify with the Fermi scale, $\Lir\sim F_{\rm{weak}}$. Between the scales $\Lir$ and $\Luv$, the dynamics is similar to walking technicolor with large anomalous dimension of the techniquark bilinear. Due to approximate scale invariance below the   scale $\Luv$  the electroweak scale remains stable \cite{Heikinheimo:2013fta}. We assume that $\Luv\lsim M_{\rm{U}}$ but allow also $\Luv\ll M_{\rm{U}}$. 
         
 In more detail, we first replace the unnatural Higgs sector of the SM with a natural strongly coupled nonsupersymmetric gauge theory whose coupling flows to a nonperturbative infrared (IR) fixed point
around the scale $\Luv$, higher than the Fermi scale $F_{\rm{weak}}$. This means that
we choose the number of technifermions, gauge group and matter representation in order to be within the conformal window \cite{Sannino:2004qp}.  We call the resulting theory {\em Untechnicolor}.  
Below the scale $\Luv$ the Untechnicolor sector becomes conformal. Once coupled to the electroweak gauge currents of the SM we expect small corrections \cite{Antipin:2013kia} which we neglect here.  

Another sector is responsible for the generation of the masses of SM fermions and pseudo Goldstone 
bosons (if present in the theory), and we consider this to be similar to extended technicolor:
There exists gauge interactions connecting ordinary matter fields and untechnicolor matter, but
these gauge interactions are broken at mass scale $M_U$. At scales below $M_U$, these interactions are summarized via the following dimension six operators:  
\begin{eqnarray}
\label{4F}
\alpha \, \frac{\overline{Q} Q \, \overline{Q} Q }{M^2_{U}} +\beta \, \frac{\overline{\psi} \psi \, \overline{Q} Q }{M^2_{U}} +\eta\, \frac{\overline{\psi} \psi \, \overline{\psi} \psi }{M^2_{U}} \ ,
\end{eqnarray}
where $\alpha$, $\beta$ and $\eta$ coefficients parametrize our ignorance of the more fundamental theory. The field $Q$ denotes an untechnicolor fermion, $\psi$ denotes a SM fermion and to keep the notation simple we have only considered one generation of SM fermions.  The $\eta$-terms may induce, depending on the underlying dynamical structure, flavor changing neutral current interactions. However, we will consider very high energy values of $M_{U}$ effectively depleting any potentially dangerous flavor changing neutral current operators.

{}To illustrate the mechanism, we consider two Dirac techniflavors $U$ and $D$ belonging to some complex representation $R$ of a generic gauge group and gauged under the electroweak group. We consider the ultraviolet (UV) operator 
${\cal O}_{UV}=\overline{U}_R U_L+\overline{D}_RD_L$ of this theory, where left- and right-handed techniquarks have usual charge assignments under the electroweak gauge group. 
Below the scale $\Luv $ the Untechnicolor sector develops a nontrivial strongly interacting IRFP described by the Lagrangian $L_{CFT}$ and due to dimensional transmutation we have 
\begin{equation}
{\cal O}_{UV}\to {\Lambda^{\gamma}_{\rm UV}} {\cal O}_{U} \ .
\end{equation} 
Here $\gamma$ is the anomalous dimension of the techniquark mass operator and ${\cal O}_U$ has dimension $d=3-\gamma$. Neglecting the small effects of the SM interactions, only the first term in Eq.\nref{4F} is relevant for the low energy effective Lagrangian  
\begin{equation}
\label{e2} 
L_{CFT}+ \alpha \frac{ {\Lambda^{2\gamma}_{\rm UV}}}{M_U^2}\,{\vert {\cal O}_{U}\vert}^{2}\,= L_{CFT}+ \frac{1}{2}\tilde{\alpha}{\vert {\cal O}_{U}\vert}^2,
\end{equation}
where $\tilde{\alpha}\equiv 2\alpha {\Lambda^{2\gamma}_{\rm UV}}/M_U^2$. This last operator can drive the theory away from conformality \cite{Sannino:2008nv} as it becomes relevant if $\gamma$ is larger than one. 
%given that it is a Nambu$-$Jona-Lasino (NJL) type of interaction. %\cite{Nambu:1961tp}. 

Without explicit mass terms, the Untechnicolor sector features a continuous mass spectrum. In order to deal with this spectrum we use the formalism introduced in \cite{Stephanov:2007ry} amounting to consider, instead of the operator ${\cal O}_U$, an infinite tower of canonically normalized massive scalar states  $\phi_{k}$,  $(k=1, 2, ..., \infty$), i.e. 
\begin{equation}
\label{e4} {\cal O}_{U}\rightarrow {\cal O} = \sum^{\infty }_{k=1}
f_{k}\,\phi_{k} \ .
\end{equation}
Here, $f^2_k(M_k^2)=\frac{B_u}{2\pi}\Delta^2(M_k^2)^{d-2}$ where
% \footnote{Generally we do not know the behaviour of $B_u$ as a function of anomalous dimension except at $\gamma=0$ and $\gamma=2$. We adapt normalization factor from \cite{Georgi:2007ek} for our quantitative estimates. See Appendix C of \cite{Sannino:2008nv} for further discussion. See also Eq. (7) of \cite{Georgi:2007ek} and Eq. (8) of \cite{Stephanov:2007ry} for the details.}
\be
\label{Georgi}
B_u=\frac{16\pi^{5/2}\Gamma(d+1/2)}{(2\pi)^{2d}\Gamma(d-1)\Gamma(2d)} \ ,
\ee
and the scalar fields $\phi_{k}$ are characterized by the mass squared $M^{2}_{k} = k\,\Delta^{2}$ 
as $\Delta\rightarrow 0$. 
Substituting this in Eq.\nref{e2} and taking into account also the mass terms of the fields $\phi_k$, the equation of motion determining the average value 
$\langle \phi_{n}\rangle$, and hence also the 
condensate $\langle {\cal O} \rangle$, reads
\begin{equation}
\tilde{\alpha} f_n \sum^{\infty }_{m=1} f_{m}\,\langle\phi_{m}\rangle=\tilde{\alpha} f_n\vev{{\cal O}}= M_n^2 \langle\phi_{n}\rangle \ .
\label{eom_condensate}
\end{equation}
Since $\vev{{\cal O}}$ is independent of $n$, this equation implies that 
%\be
$\vev{{\cal O}}\equiv {c}/\tilde{\alpha}$,
%\ee
where $c\equiv  M_n^2 \langle\phi_{n}\rangle/f_n$ is a constant.
%which implies the recursion relation $\langle\phi_n\rangle=n^{(d-4)/2}\langle\phi_1\rangle$.
%To obtain nonzero $\langle O_U\rangle$, we must have nonzero $\langle\phi_1\rangle$. 
Performing the limit $\Delta\rightarrow 0$ introduces UV and IR cutoffs defining the physical range
where the effective Untechnicolor description holds. In our case an ultraviolet cutoff is $\Luv$ since above this scale the description in terms of the composite operator ${\cal O}_U$ is no longer valid. The scale $\Lir$ is induced due to the presence of the relevant $\alpha$-coupling in Eq.\nref{e2} which breaks the conformal symmetry, and this infrared scale $\Lir$ is identified with the constituent fermion mass, i.e. the condensate
%\be
$\Lir\sim m_{\rm{const}}\sim |\vev{{\cal O}}|^{1/d}$.
%\ee
The result of the $\Delta\rightarrow 0$ limit is \cite{Sannino:2008nv}
\be
\vev{{\cal O}}=c\frac{B_u}{2\pi}\Omega(\Lir,\Luv) \ ,
\ee
where
\be
\Omega(\Lir,\Luv)\equiv \int_{\Lir^2}^{\Luv^2} dx x^{d-3} \ .
\label{OmegaFunction}
\ee
%and requiring $\langle\phi_1\rangle\neq 0$, Eq. (\ref{eom_condensate}) 

The above equations  lead to the constraint
%\footnote{Note that one obtains this same constraint from Eq. (17) in \cite{Sannino:2008nv} if $\langle O_U\rangle$ is to be nonzero in the limit $\bar{\alpha}\rightarrow 0$. Note that in Eq. (17) in \cite{Sannino:2008nv}: $b_u$ should be $b_u^2$.}
\be
\frac{c}{\tilde{\alpha}}=c\frac{B_u}{2\pi}\Omega(\Lir,\Luv) \Longrightarrow \tilde{\alpha}\frac{B_u}{2\pi}\Omega(\Lambda_{IR},\Lambda_{UV})=1 \ .
\label{constraint}
\ee
%where
%\be
%\label{OmegaFunction}
%\Omega(\Lambda_{IR},\Lambda_{UV})=\int_{\Lambda_{IR}^2}^{\Lambda_{UV}^2}dx x^{d-3}=
%\left(\frac{\Luv^{2(d-2)}-\Lambda_{IR}^{2(d-2)}}{d-2}\right).
%\ee
%The scales $\Lambda_{UV}$ and $\Lambda_{IR}$ are introduced by the limit $\Delta\rightarrow 0$ and define the domain in which the effective unparticle description is appropriate. In our case an ultraviolet cutoff is $\Luv$ since above this scale the theory flows to the UV fixed point and the description in terms of the composite operator $O_U$ is no longer valid. The scale $\Lambda_{IR}$ is induced due to the presence of the relevant $\alpha$-coupling in (\ref{e2}) which breaks the conformal symmetry. 
Using the definition of $\tilde{\alpha}$ and working in the range $1<\gamma <2$, Eq. (\ref{constraint}) leads to the relation
\begin{equation}
\label{plotted}
\Lambda_{IR}=\Luv\left[ 1+\pi \frac{\gamma-1}{B_u \alpha}\left( \frac{M_{U}}{\Luv}\right) ^2\right]^{\frac{1}{2(1-\gamma)}}.
\end{equation}
This is the advertised result of the {\it emergence} of the electroweak scale 
$\Lambda_{IR}\sim F_{\rm{weak}}$ as a result of the interplay of higher energy physical scales.

\begin{figure}[htb]
\centering
\includegraphics[width=2.5in]{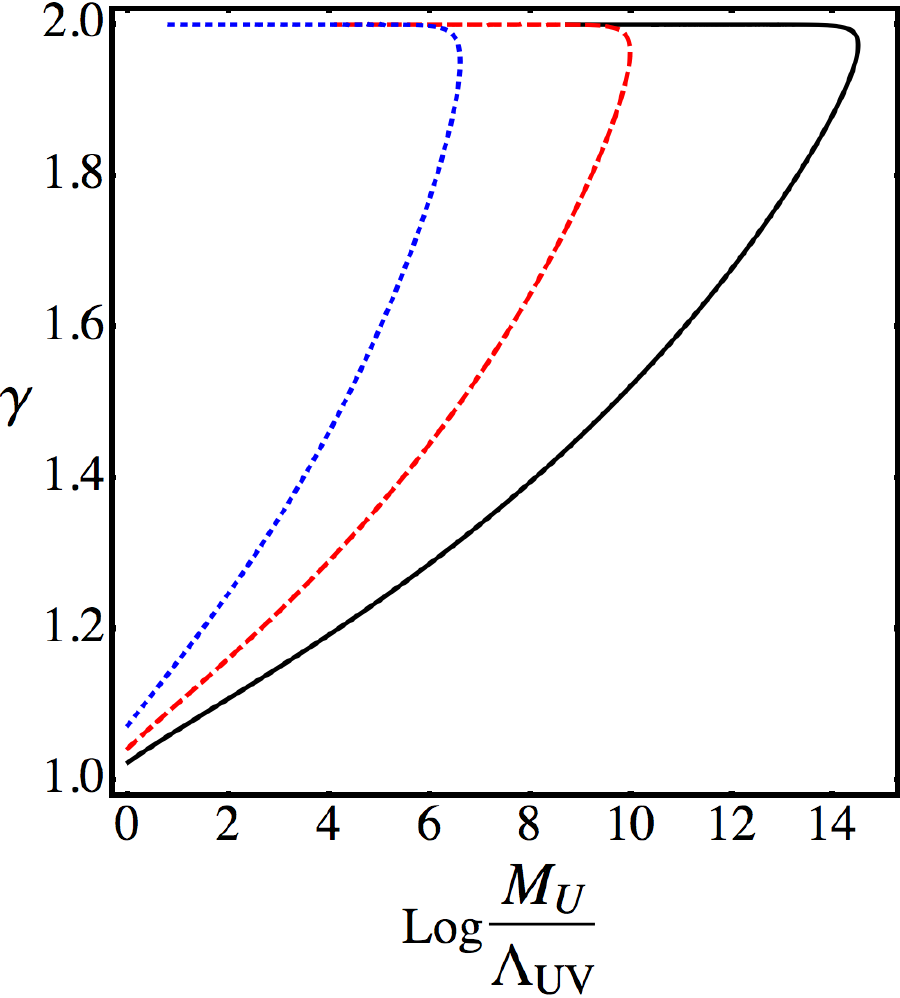} 
\caption{The curves show values of $\gamma$ and $\Luv$ which satisfy  Eq.\nref{plotted} for $\Lambda_{IR}=250$ GeV and $\alpha=0.5$.  Different curves correspond to $M_{U}=10^{16}$ GeV (solid black), $M_{U}=10^{12}$ GeV (dashed red) and $M_{U}=10^{9}$ GeV (dotted blue), }
\label{diagram}
\end{figure} 

In Fig.\ref{diagram} we plot Eq.\nref{plotted} in ($\gamma,\Luv$) plane fixing $\Lambda_{IR}=250$ GeV and $\alpha=0.5$ (we expect $\alpha$ to be of order one).  The different curves correspond to $M_{U}=10^{16}$ GeV (solid black), $M_{U}=10^{12}$ GeV (dashed red) and $M_{U}=10^{9}$ GeV (dotted blue). The electroweak scale emerges for quasi-conformal theories corresponding to these contours. 
Moreover, as Fig.\ref{diagram} implies, for given scales $\Luv$ and $M_U$, two solutions exist. One of these corresponds to $\gamma\approx2$ which mimics an elementary SM-like scalar Higgs.
%, i.e. we recover the SM Higgs sector. 
It is worth discussing this limit in more detail. Since $B_u\to 0$ as $\gamma\to 2$ (see Eq.\nref{Georgi}) and 
$\alpha$ is expected to be of order one, Eq.\nref{plotted} becomes 
\be
\label{SMHiggs}
 \quad \quad \left[\frac{\Luv}{\Lambda_{IR}}\right]^2\simeq\frac{1}{B_u}\left(\frac{M_{U}}{\Luv}\right)^2,
 \qquad (\gamma\simeq 2).
\ee
 Having assumed  $\Luv \ll M_{U} $ and without fine-tuning $B_u$ we observe that the hierarchy between $\Lir$ and $\Luv$
Eq.\nref{SMHiggs}  derives from a seesaw-like mechanism   
$\Lambda_{IR} \sim  {\Lambda^{2}_{\rm UV}} /M_U$. 
In Fig.\ref{diagram}, when approaching $\gamma=2$ from below along a curve corresponding to fixed values of $M_{\rm{U}}$ and $\Lir$,
this scaling relation corresponds to the reach of the plateau at  the maximum possible value of $M_{\rm{U}}/\Luv$. For example, for the solid black curve corresponding to $M_{U}=10^{16}$ GeV this happens for 
$\log(M_{\rm{U}}/\Luv)\sim \log(M_{\rm{U}}/\Lir)^{1/2} \approx 15$ which translates into $\Luv\sim 3 \times 10^{9}$ GeV. At smaller values of $M_{\rm{U}}/\Luv$, the plateau extends to the region $\Luv \approx M_U$ with increased tuning of the coefficient $B_u\to (\Lir/\Luv)^2$.

%value $\rm{Log}\sqrt{M_U/\Lir}$ as $\gamma$ increases from 1 to 2. For example, for the solid black curve corresponding to $M_{U}=10^{16}$ GeV this happens at  $\approx \rm{Log}\sqrt{10^{13}}\approx 15$ which translates into $\Luv\sim 3 \times 10^{9}$ GeV. After that, the plateau extends to the region $\Luv \approx M_U$ at the price of the extreme tuning of the coefficient $B_u\to \Lir/\Luv$.

At the lower limit, $\gamma\approx 1$, Eq.\nref{OmegaFunction} reduces to 
$\Omega(\Lambda_{IR},\Lambda_{UV})=\log{(\Luv/\Lir)^2}$ and therefore the constraint in Eq.\nref{constraint} translates to 
%($\Lambda_{UV}=\Luv$)
\be
\Lambda_{IR}=\Luv \exp \left[-\frac{\pi}{\alpha B_u}\right] \ ,
\ee
where we have replaced $\tilde{\alpha}$ with $\alpha$ because, as is clear from Fig.\ref{diagram}, $\Luv\approx M_U$ in this limit and therefore $\tilde{\alpha}\approx \alpha$. We thus see that in this limit the model features only one high energy scale, $\Luv$, and
% which is compatible with the definition for the conformal phase transition in the sense that the 
the dynamically generated infrared scale $\Lambda_{IR}$ is exponentially suppressed with respect to 
$\Luv$. 

%\section{Discussion}
We have considered a model where all beyond SM dynamics occurs at very high scales and the EW symmetry breaking, hence the Fermi scale itself, emerges as a low energy mirage. In addition to giving a novel explanation for the existence of the Fermi scale, the model also provides further phenomenological implications which we will now discuss. Let us start with the top mass. In the limit $\gamma\approx 2$ and $M_{\rm{U}}\approx \Luv$ it can be generated with the $\beta$-term in Eq.\nref{4F}:
\be
\frac{\overline{t} t \, \overline{Q} Q}{M_U^2} \quad \Longrightarrow \quad m_t \sim  \frac{\langle{\cal O}_{\rm{U}}\rangle_{M_U}}{M_U^2},
\ee
where $\langle{\cal O}_{\rm{U}}\rangle_{M_U}$ condensate is evaluated at the scale $M_U$. The large value of the anomalous dimension enhances the value of the condensate at high energy and for $\gamma=2$ we have
\be
\frac{\langle{\cal O}_{\rm{U}}\rangle_{M_U}}{\langle{\cal O}_{\rm{U}}\rangle_{\Lir}}=\bigg(\frac{M_U}{\Lir}\bigg)^{\gamma=2} \ .
\ee
Using that $\langle{\cal O}_{\rm{U}}\rangle_{\Lir}\sim \Lambda^3_{IR}$ we obtain $m_t\sim \Lir$ in agreement with experiments.

Generating the large top mass becomes progressively more difficult when the hierarchy between $\Luv$ and $ M_{\rm{U}}$ grows. Assuming QCD-like running behavior between scales $M_U$ and $\Luv$ leads to 
$\langle{\cal O}_{\rm{U}}\rangle_{M_U}\approx \langle{\cal O}_{\rm{U}}\rangle_{\Luv}$ and therefore
\be
m_t \sim \bigg(\frac{\Luv}{M_U}\bigg)^{2} \Lir  \  .
\ee

Thus, additional contributions are needed below the flavor scale $M_U$ in order to bring the top quark mass to its experimental value. To extend the present context we may imagine a conformal topcolor scenario where relevant perturbations of the scale invariance at $\Luv$ would lead to both the electroweak scale and the top mass itself dynamically at low energies. We do not attempt a detailed solution of this idea here. 

The phenomenology of these kind of models  is  similar to the one of  ideal walking technicolor \cite{Fukano:2010yv,Dietrich:2006cm} where chiral symmetry breaking (and therefore conformality) is driven by four-fermion operators. In this scenario, if the conformal transition is walking and not jumping \cite{Sannino:2012wy,Antipin:2012sm}, we expect the models to feature a compressed tower of composite states at the induced electroweak scale, \cite{Catterall:2007yx,Hietanen:2008mr,Bursa:2011ru,Fodor:2012ty}. 
For behaviors at finite temperature, see \cite{Tuominen:2012qu}.
The collider phenomenology of these models, expected to carry over to the present idea, has been investigated in \cite{Foadi:2007ue,Belyaev:2008yj}. Another important issue that any dynamical mechanism faces is how to obtain naturally (i.e. without invoking {\it special} dynamics) the correct physical mass of the composite Higgs state. The answer to this important point has been recently put forward in \cite{Foadi:2012bb} and relies on the fact that the observed physical mass of the composite Higgs is due to the interplay of the composite dynamics and the corrections due to the coupling to the top (yet another four-fermion induced operator) which tends to lower the composite Higgs mass towards the observed value. Disentangling ideal walking from the models put forward here (technically an extreme case of ideal walking)  requires stronger constraints on the new flavour physics scale and the knowledge of the detailed spectrum at the electroweak scale. 

{Till now we assumed that the gauge dynamics and the associated conformal breaking occurred via natural theories, i.e. gauge theories with only fermionic fundamental matter fields. The point we have proven is that the interplay of several natural fundamental sectors at very high energy can lead to the existence of the substantially lower electroweak scale.  

Before concluding we would like also to speculate on the effects of the introduction of further irrelevant operators on near conformal dynamics. These operators, as we shall argue, might be useful for the generation of neutrino masses. To illustrate the mechanism we consider a toy construction making use of fundamental scalars, although the scenario could be later replaced by a more fundamental dynamics.} Consider a SM-singlet complex scalar field $S$ and further assume that $S$ acquires a vacuum expectation value (vev) $\Sigma$. The lowest dimension-8 operator coupling $S$ with ${\cal O}_{\rm{UV}}$ 
is the following:
\be
\kappa\frac{|S|^2|{\cal O}_{\rm UV}|^2}{M_{\rm{U}}^4}, 
\ee
where $\kappa$ is a dimensionless coupling constant of $O(1)$.
After the SM-singlet scalar $S$ acquires a vev, we arrive at the following effective Lagrangian density
\begin{equation}
\label{e21} 
L_{CFT}+ \frac{1}{2}
\tilde{\alpha}_{\rm{eff}}{\vert O_{U}\vert}^{2},
%-\delta \Sigma^2 \sum^{\infty }_{n=1} \phi_{n}^2
\end{equation}
where $\tilde{\alpha}_{\rm{eff}}\equiv 2(\alpha+ \kappa\frac{\Sigma^2}{M_U^2})\frac{\Lambda_{UV}^{2\gamma}}{M_U^2}$ and $\alpha$ and $\kappa$ are dimensionless couplings expected to be of $O(1)$. For $\kappa=0$ we recover Eq.\nref{e2}.
%Compared to the setup of the previous section we added the $\delta$-coupling \cite{Delgado:2007dx} which respects the conformal symmetry but cannot be expressed in terms of $O_U$ in the continuum limit. 

The Lagrangian of Eq.\nref{e21} has the same form as Eq.\nref{e2} with $\tilde{\alpha} \to \tilde{\alpha}_{\rm{eff}}$ and, for $\kappa >0$, $\tilde{\alpha}_{\rm{eff}}> \tilde{\alpha}$. The addition of the complex scalar therefore increases the strength of the operator that drives the theory away from conformality. %and is important if the coupling strength in the absence of this effect is not large enough to break the conformal symmetry. 
For this effect to be non-negligible it must be that $\langle S\rangle= \Sigma \approx M_U$. 

Within this setting now, consider the Yukawa sector for the neutrino sector of the SM, introducing also right handed neutrinos:
\be
{\cal L}_{\rm{Yuk.\nu}}=-y L_L{\cal O}_U\nu_R+\lambda S\bar{\nu}^c_R\nu_R+{\textrm{h.c.}} \ ,
\ee
where $L_L$ is the usual SM lepton doublet and $\nu_R$ is the right handed neutrino, $S$ and ${\cal O}_U$ are the SM singlet scalar and low energy composite Higgs fields. The usual Yukawa coupling leading to Dirac mass is denoted by $y$ and the coupling leading to Majorana mass is $\lambda$. As the singlet $S$ condenses, a Majorana mass $M=\lambda \Sigma\sim M_U$ is generated, which is much greater than the value of the condensate $\langle{\cal O}_U\rangle\sim  {\Lambda^{2}_{\rm UV}}/M_U\sim F_{weak}$ which generates a Dirac mass $m_D\sim y F_{weak}$. Therefore we obtain the conventional type-I seesaw mass matrix for the neutrinos which gives two physical Majorana eigenstates. First, a superheavy state with mass $m_1\sim M_U$ and an extremely light state with mass
\be
m_2\sim \frac{y^2 {\Lambda^{4}_{\rm UV}}}{\lambda {M_U^3}}\sim \frac{y^2}{\lambda}\frac{F_{weak}^2}{M_U} \ .
\ee
Hence, we can explain the existence of the hierarchical scales observed in nature: the high scale $M_U\sim  10^{16...19}$ GeV, which sources the breaking of the scale invariance of the postulated new physics underlying the observed scalar sector of the SM, the electroweak scale $F_{weak}=246$ GeV and the sub-eV scale of neutrino masses $m\sim 10^{-2}$ eV. For alternative model setups of this type, see e.g. \cite{Atwood:2004mj}. Among the new features is the existence and use of an intermediate scale $\Luv\sim 10^{10}$ GeV, also implied from the stability analysis of the SM with the observed Higgs sector. 
 
To summarize, we have considered the possibility that the electroweak scale emerges via the interplay 
between two higher energy physical scales. As a concrete example, we considered a model framework consisting of a theory possessing a nontrivial IRFP at scales below $\Luv$ but perturbed 
by a four fermion coupling. We assumed that such coupling originates from a more complete theory above the scale $M_{\rm{U}}\gsim\Luv$. The resulting dynamics below the scale $\Luv$ is unparticle-like, and we termed it Untechnicolor. Assuming that the value of the four-fermion coupling is sufficiently large to enforce the dynamical vacuum expectation value for the scalar Untechnicolor operator, we demonstrated how the electroweak scale, i.e. $246$~GeV, identified with $\Lir$ arises. Below the electroweak scale the Untechnicolor sector turns into an effective SM Higgs-like sector. 

Our construction explains how the electroweak scale arises due to small explicit breaking of scale invariance and how the electroweak scale is stable under radiative corrections due to approximate scale invariance below the compositeness scale $\Luv$. Furthermore, in the scenario we considered, the flavor physics scale can be around the compositeness scale $\Luv$ or significantly above it. 
%We found that this possibility exists for the class of theories summarized numerically on Fig. \ref{diagram}.

\begin{acknowledgments}
This work is partially supported by the Danish National Research Foundation under the contract number DNRF90. Support from the European Science Foundation (ESF) within the framework of the ESF activity entitled 'Holographic Methods for Strongly Coupled Systems' is gratefully acknowledged.
\end{acknowledgments}

\end{document}